# Magnetoplasmonics: current challenges and future opportunities

Nicolò Maccaferri[1,2], Alessio Gabbani[3], Francesco Pineider[3], Terunori Kaihara[4], Tilaike Tapani[1], Paolo Vavassori[4,5]

[1]Department of Physics, Umeå University, Linneaus väg 24, 901 87, Umeå, Sweden
[2]Department of Physics and Materials Science, University of Luxembourg, 162a avenue de la Faïencerie, L-1511 Luxembourg, Luxembourg
[3]Department of Chemistry and Industrial Chemistry, University of Pisa, Via Moruzzi 13, 56124 Pisa, Italy
[4]CIC nanoGUNE BRTA, Tolosa Hiribidea, 76, 20018 Donostia-San Sebastián, Spain
[5]IKERBASQUE, Basque Foundation for Science, Plaza Euskadi 5, 48009 Bilbao, Spain

**Plasmonics represents a unique approach to confine and enhance electromagnetic radiation well below the diffraction limit, bringing a huge potential for novel applications, for instance in energy harvesting, optoelectronics, and nanoscale biochemistry. To achieve novel functionalities, the combination of plasmonic properties with other material functions has become increasingly attractive. In this Perspective, we review the current state of the art, challenges, and future opportunities in nanoscale magnetoplasmonics, an emerging area aiming to merge magnetism and plasmonics in confined geometries to control either plasmons, electromagnetic-induced collective electronic excitations, using magnetic properties, or magnetic phenomena with plasmons. We begin by highlighting the cornerstones of the history and principles of this research field. We then provide our vision of its future development by showcasing raising research directions in mid-infrared light-driven spintronics and novel materials for magnetoplasmonics, such as transparent conductive oxides and hyperbolic metamaterials. As well, we provide an overview of recent developments in plasmon-driven magnetization dynamics, nanoscale optomagnetism and acousto-magnetoplasmonics. We conclude by giving our personal vision of the future of this thriving research field.**

## Introduction

The ability to control materials at the nanoscale allows to confine light into subwavelength volumes by exploiting collective electromagnetic-induced electronic excitations, known as plasmons.[1–6] Unlike conventional optics, plasmonics enables unrivaled confinement and enhancement of the electromagnetic (EM) radiation well below the diffraction limit, containing a huge potential for real-life applications in energy harvesting, wave-guiding and lasing, optoelectronics and nanoscale biochemistry.[7–9] To achieve novel functionalities, the combination of plasmonic properties with other material functions has become increasingly appealing. In the past two decades nanoscale magnetoplasmonics has been an emerging area aiming to merge magnetism and plasmonics in confined geometries to control in new ways either the properties of plasmons using magnetic properties of matter or magnetic phenomena with plasmons.[10–14] In this Perspective, we review the current state of the art, challenges, and future opportunities in some thriving areas of magnetoplasmonics. We begin with the use of plasmons to enhance magneto-optical (MO) effects in metallic systems, in the framework of active flat-optics metamaterials for light polarization control, lasing, nonlinear magneto-optics, as well as for biochemical and chemical sensing, highlighting its history cornerstones and main principles and providing our vision of its future development.

We then venture into the new trends in the field. Relevant examples are mid-infrared MO spintronics, the exploration of new and lossless materials, such as conductive oxides, dielectrics, and semiconductors, as well as near-zero index and hyperbolic metamaterials. We will also briefly survey the latest advances in plasmon-driven magnetization dynamics and optomagnetism, e.g., the enhancement of the inverse Faraday effect in nanoscale geometries and the use of orbital angular momentum of light to achieve a superior ultrafast control of magnetism and spintronics.

We conclude giving our outlook on the future of the field, showcasing new possible directions to achieve a full control of MO effects and their enhancement by using nanoscale materials, as well as drive magnetic phenomena with plasmons at the atomic and sub-femtosecond timescales.

## Magnetoplasmonics in confined geometries

The recent development of nanophotonics has allowed for active and flexible light control and enabled numerous technological applications such as metasurface polarization switches,[15,16] chirality-sensitive nanoantennas,[17] high resolution imaging and high sensitivity molecular detectors,[18–20] as well as single-photon integrated quantum circuits.[21] The MO effects can offer the active and nonreciprocal control of the polarization and intensity of reflected and transmitted light. Therefore, the integration of MO-active materials in nanophotonics is particularly appealing for a variety of applications: ultrathin devices for optical isolation and modulation, information processing and cryptography, sensing, imaging, structured light and holography, etc.

Unfortunately, MO materials do not display sufficiently large MO effects when downsized to a sub-micron scale (e.g., Faraday rotation of a 1-µm thick Ce:YIG is only 0.5° in the infrared[22]), and this has hindered the utilization of magnetoplasmonics in, for instance, active nanophotonics or in light-driven nanomagnetism and spintronics, where a major goal is a coherent control of nm-sized magnetic bits.[14,23]

Magnetoplasmonics utilizes the peculiarities of optics at the nanoscale to enhance MO effects through the excitation of localized surface plasmons (LSPs) in metallic magnetic nanomaterials. Several-fold enhancement of magnetoplasmonic effects was indeed reported in nanostructures composed of pure ferromagnetic (FM) nanoantennas,[24–31] while noble metal nanostructures display 2 orders of magnitude weaker magnetoplasmonic response at magnetic fields of the order of the Tesla.[32–35] Even so, the LSP-mediated enhancement of MO is not enough to enable practical applications. The only area in which currently magnetoplasmonics can be advantageously applied is molecular and refractive index sensing.[27,33,36–40]

According to a semi-classical theoretical description,[29] the enhancement of the MO activity in a magnetoplasmonic nanoantenna resonator, compared to a film of the same material and the same thickness, is understood as due to a second spin-orbit coupling induced LSP (magneto-optical LSP, MO-LSP) driven by the LSP directly excited by the incident linearly polarized light. These two orthogonal LSPs display a phase lag and different amplitudes (see Figure 1). The far-field MO response of the magnetoplasmonic nanoantenna results then from the ratio between the electric dipoles associated to the induced SOC-LSP and the directly excited LSP. In the case of a cylindrical magnetoplasmonic nanoantenna, the amplitudes of the LSP and MO-LSP electric dipoles are proportional to Q and $Q^2$, respectively, where Q is the quality factor of the resonance. Ultimately, the MO activity of a magnetoplasmonic nanoantenna is Q-times larger than that without LSP (e.g., that of a continuous film of the same thickness and material of the nanoantenna).

The quality factor Q is governed by the absorption and radiation losses, and different strategies have been explored to reduce such losses and achieve high values of Q. Absorption losses are particularly high in FMs, which result in a very small Q for pure FM nanoantennas (in the order of 2-5 to be compared to values in the order of 10 for noble metals in the same spectral range). Therefore, much effort has been devoted in devising hybrid structures made of FM and noble metals/non-magnetic (NM) materials, which chiefly contribute to the reduction of absorption losses bringing the Q-factor closer to that of a pure noble metal nanoantenna.[41–47] Interestingly, such hybrid systems have been recently reported to enable lasing effects, which can be also tunable by using external magnetic fields.[48,49]

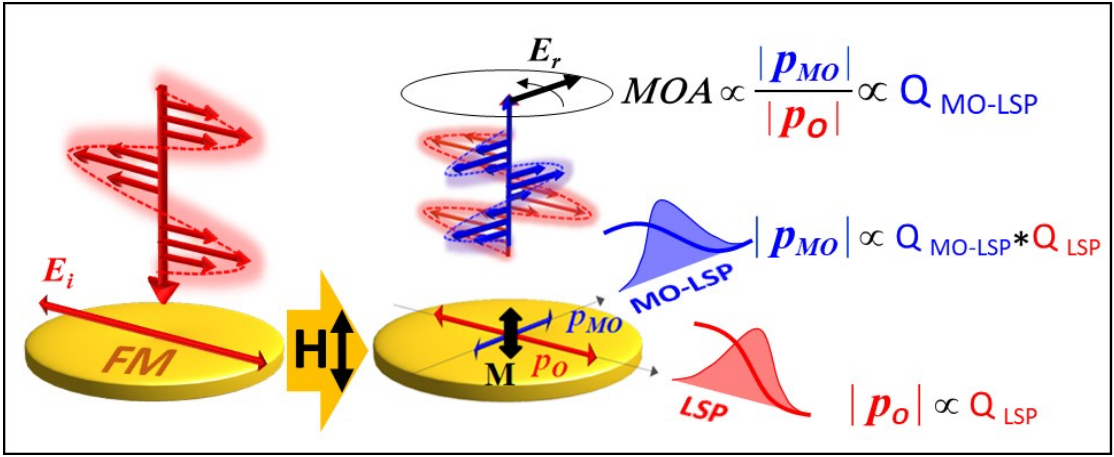

**Figure 1. Semi-classical model describing the interaction of linearly polarized light with a magnetoplasmonic nanoantenna.** The figure sketches the electrodynamics of the ferromagnetic (FM) disk generating an electric dipole ($p_O$) resonantly triggered by the electric field $E_i$ of an incident linearly polarized electromagnetic radiation and a magneto-optically activated (via spin-orbit coupling) electric dipole ($p_{MO}$) by the application of a magnetic field H to induce a magnetization M in the nanoantenna as shown. The enhancement of the amplitude of $p_O$ at resonance, is proportional to the Q-factor of the resonance along the direction of $E_i$ ($Q_{LSP}$), while that of $p_{MO}$, induced by $p_O$ and orthogonal to it, is proportional to the product of the $Q_{LSP}$ and the Q-factor of the resonance in the orthogonal direction ($Q_{MO\text{-}LSP}$). The enhancement of the amplitude of the polarization change induced in the re-emitted electromagnetic radiation (magneto-optical activity, MOA) is proportional to the ratio between the module of $p_{MO}$ and $p_O$ and thereby limited to a factor $Q_{MO\text{-}LSP}$ despite the much larger enhancement of $p_{MO}$.

The suppression of radiation losses in specific frequency ranges can be achieved by arranging plasmonic nanoantennas in a periodic array thanks to the diffracted coupling of the EM field radiated by each nanoantenna. Diffractively coupled LSP is referred to as plasmonic surface lattice resonance (SLR),[50–53] and its excitation contributes to the increase of the Q-factor of the collective resonance of the array. By employing plasmonic SLRs in magnetoplasmonic arrays (viz., magnetoplasmonic crystals), a rich variety of interference features and further enhancement of MO activity have been reported [45,54–57].

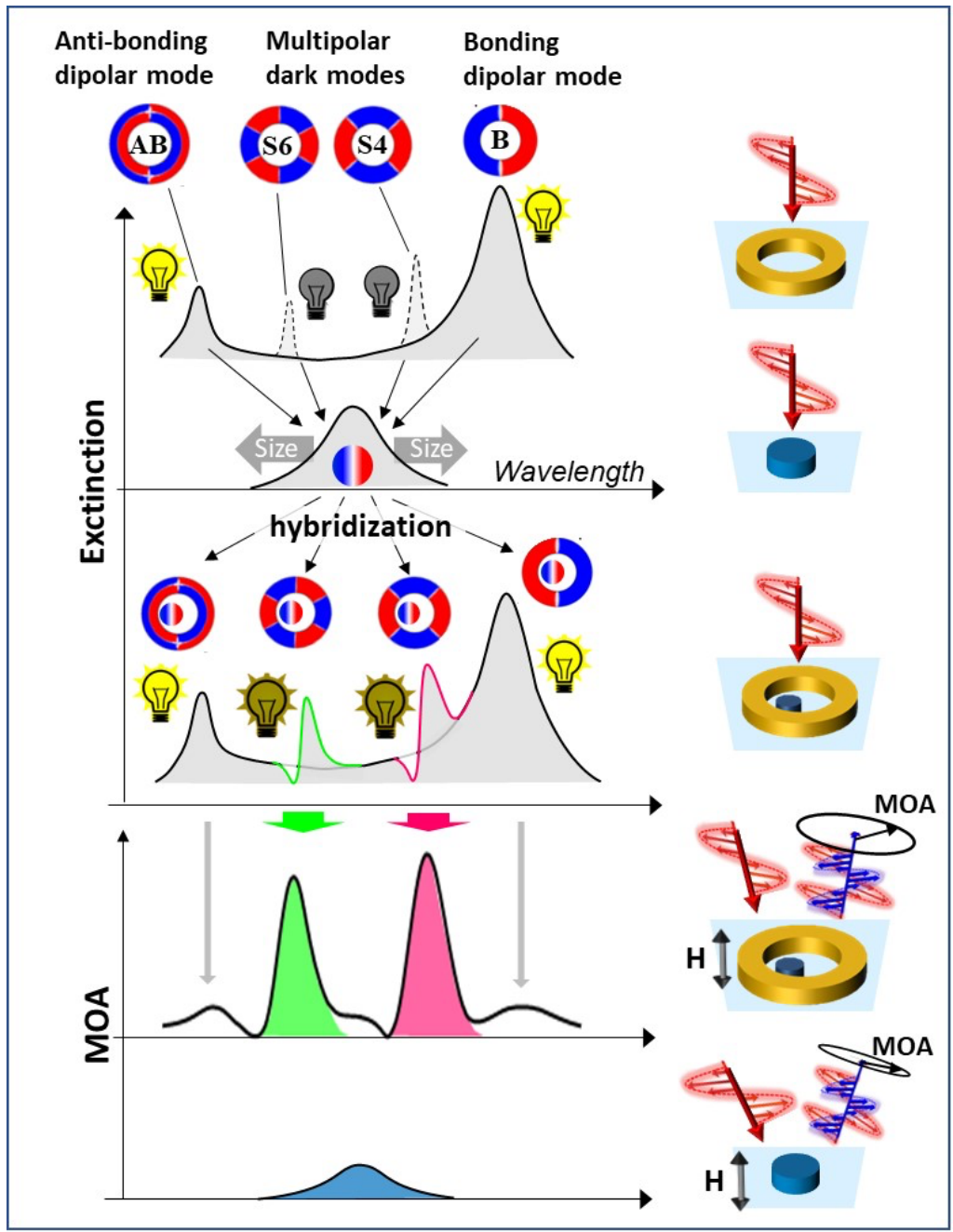

**Figure 2. Beyond the typical concept of LSP to enhance the MO response in confined geometries.** General concept of hybridization with dark modes to boost the magneto-optical response, illustrated utilizing the hybrid ring-dot structure reported in Ref. [58]. The hybridization between the bright dipolar mode of the ferromagnetic dot with multipolar dark modes of a surrounding noble-metal ring results in Fano-like low radiant modes. The case of the hybridization with the hexapolar (S6) and quadrupolar (S4) modes is shown as an example. The excitation of one of these hybrid modes results in a strongly enhanced bright magneto-optical dipole in the ferromagnetic dot. The EM radiation with the incident polarization is re-emitted by the low-radiant hybrid Fano mode. The EM radiation with the polarization rotated by 90º is radiated by the enhanced bright magneto-optical dipole. Thereby, the magneto-optical activity (MOA), which results from the ratio of the two re-emitted radiation, results strongly enhanced with respect to that achievable with a magneto-plasmonic antenna (e.g., a 10-fold enhancement is reported in Ref. [58] compared to the results obtained with pure Ni disk nanoantennas in Ref. [29]).

Despite all the efforts devoted to increase Q, the achieved amplification of the MO effect is still not enough to enable applications. It is worth noting that the SOC-LSP electric dipole is proportional to $Q^2$, which, upon reduction of absorption and radiation losses applying the strategies described above, can be of the order of 100 or even higher.[58] The fact that the MO response is proportional to the ratio of the MO-LSP and LSP electric dipoles limits its enhancement only to a factor Q, and thus limited to 10-20 even upon reduction of losses. Physically, this can be understood as due to the simultaneous amplification of the re-emitted light by the LSP with the incident polarization. If the enhancement of the re-emission of light with the original polarization could be contained, amplifications of the MO effect up to $Q^2$ could be achieved.

A recent study has shed light on this hidden amplification potential.[58] It relied on non-concentric nanocavities consisting of a NM ring resonator near-field coupled to a FM disk. The near-field coupling induces the hybridization of the electric dipole of the FM disk with a multi-polar dark mode of the surrounding NM ring, leading to a Fano-like resonance for the hybrid mode.[59,60] The concept is illustrated in Figure 2. The result of the coupling leads to two key effects: i) the electric dipole induced in the FM disk is even larger than the LSP of the isolated disk; ii) the multipolar mode of the NM ring acquires an electric dipole component that is oscillating out of phase with that induced by the incident light in the FM disk. The combination of these two effects benefits the MO activity enhancement in two ways: effect i) corresponds to an increase of the LSP dipole amplitude, and thus of the SOC-LSP, beyond the Q value of the FM disk alone; effect ii) allows for the management, in particular the reduction, of the radiated light with the original polarization of the incident field, thereby enabling a MO enhancement larger than Q that can eventually approach the theoretical $Q^2$ limit. For instance, the magnetoplasmonic nanocavity studied in Ref. [58] exhibited 7-fold enhancement in MO activity compared with that produced by the excitation of a LSP in the bare FM disks.

**New directions from a material point of view**

Enabling a MO enhancement larger than Q seems very promising and further research should consider the followings: low dissipative materials (NM/FM hybrids,[46,58] dielectrics,[61,62] conductive oxides,[63] etc.), small cavity footprints (higher surface coverage and MO effects), functional arraying (SLR,[53] metagratings,[64] etc.), anisotropic superlattice films (Au/Co, Au/Fe, Pt/Co, etc.[65]), epsilon-near-zero (ENZ) and near-zero-index (NZI) materials,[66–68] and hyperbolic metamaterials.[69–72] Several efforts in the design of NM/FM hybrids have been reported by designing hybrid nanostructures made of noble metals (Au or Ag) and iron oxides or magnetic metals. Coating Au nanospheres (8 nm) with a thin $FeO_x$ shell (1 nm) was found to enhance the MO response up to 50% with respect to uncoated Au nanospheres, which was ascribed to a purely dielectric effect.[73] However, if the volume of the iron oxide shell is increased, the MO response if fully dominated by the MO signal of the iron oxide moiety, masking any magnetoplasmonic effect.[74,75] On the other hand, improvement of the magnetoplasmonic response was observed in different combination of materials, such as Ag@FeCo,[76] and $FeO_x$@Au[77] NPs prepared by chemical methods or sandwich Au/Co/Au nanodisks prepared by lithographic techniques.[78,79] However, the enhancement of the MO response is not enough to exceed the one of ferromagnetic nanodisks, which remain the state-of-the art material for nanoscale magnetoplasmonics. While strategies based on NM/FM hybrids and dielectrics have been widely reported and studied in the past decade,[12] here we would like to highlight two emergent approaches which can contribute to fuel the development of future exciting new directions in this thriving research field: conductive oxides and hyperbolic metamaterials.

As mentioned in the previous section, critical limitations for the enhancement of the magnetoplasmonic response from the material point of view depend on the Q-factor of the LSP. However, these arguments are valid only for standard metals, either pure noble or FM. In recent years several new plasmonic materials have emerged, enriching the field of plasmonics and photonics: dielectric antennas[61,62,80] and degenerately doped nanostructures[81–85] are among the most promising systems, but have been scarcely explored for magnetoplasmonics, thus holding an unexploited potential. Among doped semiconductors, transparent conductive oxides (TCOs) represent a valuable material choice for magnetoplasmonics. In NM plasmonic nanoparticles the magnitude of the magnetic modulation is proportional to the cyclotron frequency ($\omega_c=qB/m$), which is inversely proportional to the effective mass of charge carriers ($m$). While in metals $m$ is fixed to values close to the free electron mass $m_e$, in TCOs this value is typically much lower (0.2-0.3$m_e$). Moreover, typically conductive oxides such as ITO nanocrystals (NCs) display lower carrier densities than metals, shifting the plasmonic resonance in the near infrared, which is typically exploited for several applications spanning from thermoplasmonics-based technologies[86,87] to enhanced infrared spectroscopy.[88] Carrier charge can also vary if the carriers are either n- (electrons) or p-type (holes or oxygen vacancies): changing the sign of the charge of the carriers induces an opposite sign of the MO ellipticity.[89] Interestingly, in colloidal TCO NCs, the damping parameters can be considerably reduced through dopant choice and defect engineering, allowing to reach Q-factors values up to 12 (which is close to noble metals despite the 2 order of magnitudes lower carrier densities) in simple 10-20 nm spherical nanoparticles supporting electric dipole modes.[90–92] Radiative damping due to scattering is strongly reduced in TCO nanostructures having LSP above 1700 nm, as the scattering cross section for spheres is proportional to $\lambda^{-6}$ and is thus strongly reduced at such long wavelength (by approximately 3 orders of magnitude compared to nanostructures of the same volume having the LSP at 550 nm).[93] A recent work demonstrated that colloidal TCO NCs enable to reach quite large magnetoplasmonic response, challenging metallic nanostructures, due to the large Q-factor combined with the large cyclotron frequency (see Figure 3).[63] The importance of keeping the plasmonic resonance sharp is highlighted by the larger magnetoplasmonic response in FICO (F$^-$ and In$^{3+}$ doped CdO) with respect to ITO (Sn-doped In$_2$O$_3$). Proof of concept refractometric sensing experiments demonstrated the higher performances of TCO NCs also compared to ferromagnetic materials (Ni nanodisks). However, such performance needed higher applied magnetic fields (around 1.4 T) with respect to Nickel (0.2-0.3 T). While such applied field can be easily achieved with hard disk writing heads,[94] the possibility to work at lower applied field would be preferable for magnetoplasmonic sensing devices in terms of economic costs, and size of the final device.

To this purpose, the magnetoplasmonic response of TCOs can be potentially further improved by introducing magnetic cations as co-dopants. Indeed, several reports of carrier-mediated interaction between magnetic dopants were reported in the literature of dilute magnetic semiconductors, in some cases also reaching ferromagnetism at room temperature.[95] A clear example is represented by ZnO films doped with Co$^{2+}$ or Mn$^{2+}$ in the presence of free carriers.[96,97] Using nanostructures of these materials and tailoring their geometric and electronic features in order to have a high Q-factor, could potentially increase the coupling of free charge carriers with external magnetic fields, finally improving the magnetoplasmonic response of TCO nanostructures.

Recently, interesting correlations between localized plasmons and the MO response at the excitonic resonance have also been reported in doped semiconductor nanocrystals by Radovanovic *et al.*, where non-resonant coupling between excitons and the circular magnetoplasmonic modes has been observed. Such effect results in the observation of the Zeeman splitting of the excitonic states in degenerately doped In$_2$O$_3$ NCs, which is not observed in undoped NCs (where no carrier density is present in the conduction band), leading to a robust electron polarization at room temperature, opening interesting implications for spintronics and quantum information technologies.[98–100]

Remarkably the large degree of tunablity of charge carrier parameters (mass, charge and density of the carriers) in these materials offers a great opportunity for boosting magnetoplasmonic devices performance close to those needed for active nanophotonic devices.

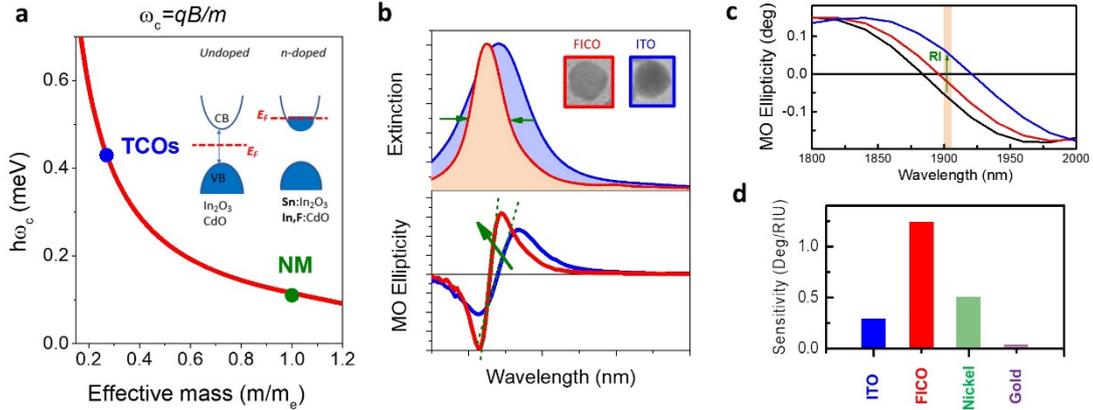

**Figure 3. Magnetoplasmonics beyond metals: transparent conductive oxides.** a) The cyclotron frequency ($\omega_c$) is plotted as a function of the effective mass of charge carriers to show the main advantage of TCOs over noble metals (NM) for magnetoplasmonics. In the inset, the typical effect of aliovalent doping on the band structure of the TCO semiconductor is shown. b) Extinction spectra and Faraday ellipticity of Sn-doped $In_2O_3$ (ITO) and F- and In-doped CdO (FICO) NCs, showing the effect of the reduced plasmon line width of FICO on the MO response, increasing the slope of the MO ellipticity close to the plasmonic resonance. c) MO ellipticity measured at 1.4 T in three solvents with different refractive index ($CCl_4$, $C_2Cl_4$ and $CS_2$). d) The MO refractive index sensitivity of ITO and FICO NCs, defined as MO ellipticity variation per refractive index unit at fixed wavelength is compared with nickel nanodisks and gold nanoparticles.[63]

It is worth mentioning that ENZ and NZI materials, which display a phase refractive index close to zero at specific wavelengths, are known to enhance or inhibit light-matter interactions, thus representing a new frontier in magneplasmonics and magneto-optics. In fact, they can, for instance, allow the excitation of additional plasmonic/photonic modes (such as strong and sharp bright magnetic modes) in confined nm-size geometries which can be combined with the previously reported strategies to enhance Q beyond the current limits. Thus, they represent an additional and very interesting playground to explore new ways for enhancing MO activity at the nanoscale.

A concrete example of this potential is represented by a sub-class of these materials, called hyperbolic metamaterials (HMM). One of the most interesting properties of these materials is that they display a strong optical anisotropy and nontrivial properties, such as conductive behavior along particular spatial directions and insulating behavior along the others, even in nanoscale geometries like antennas or plasmonic crystals.[69,70,102–106]

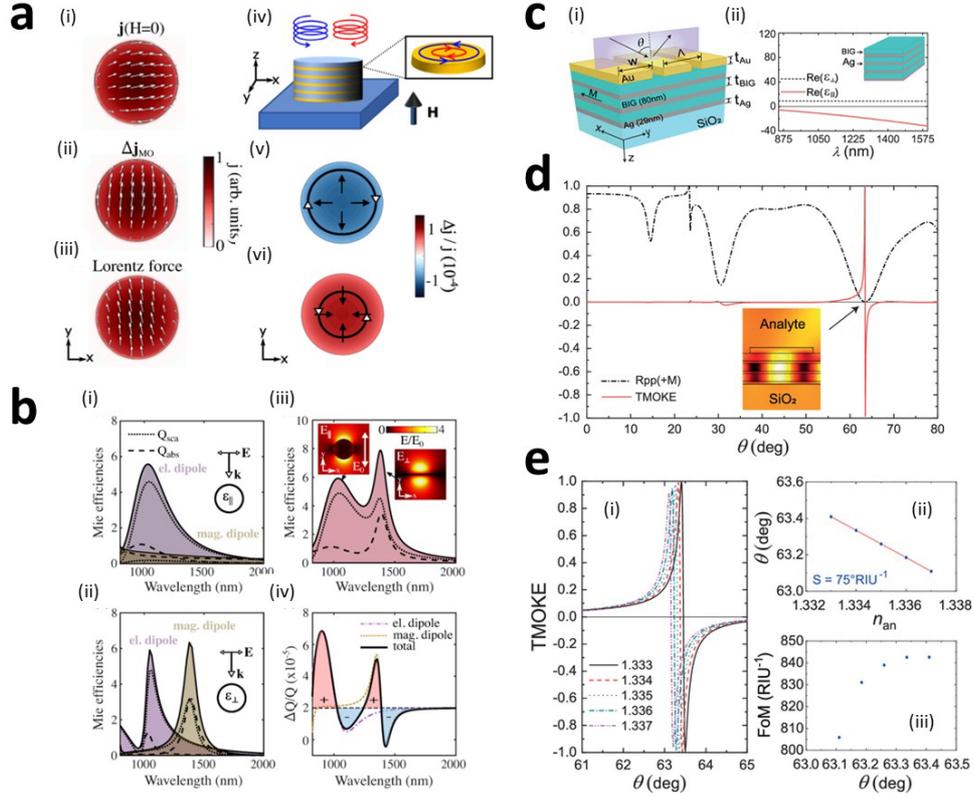

**Figure 4. Magnetoplasmonics beyond metals: hyperbolic metamaterials.** a) (i)–(iii) Calculated current density **j** (white arrows) at the low-energy resonance for zero magnetic field (i), magnetically induced change of current $\Delta \mathbf{j}_{MO}$ (ii), and Lorentz force calculated as cross product of **j** and the unit vector in direction of **H** (iii). The absolute value of **j** is color coded for all three cases. (iv) Circular dichroism effect in terms of spatial confinement or broadening of circular modes. (v),(vi) Induced relative change of current density $\Delta \mathbf{j}/\mathbf{j}$ (color coded) for opposite helicity of the incoming electric field, as sketched in the plot. Black arrows show the gradient of $\Delta \mathbf{j}/\mathbf{j}$ and display the induced spatial broadening or confinement of the circular plasmonic motion. b) Scattering (dotted lines) and absorption (dashed lines) efficiencies $Q_{sca}$ and $Q_{abs}$ for spherical nanoparticles with isotropic permittivity $\epsilon_{\parallel}$ (i) and $\epsilon_{\perp}$ (ii). Interaction efficiencies are analytically calculated with the Mie expansion up to the first order, where magnetic (green) and electric (purple) contributions are separately evaluated. (iii) Sum of the in-plane electric dipole shown in (i) and the out-of-plane magnetic dipole shown in (ii). Inset are near-field plots of the resonances normalized to the incoming field $E_0$ indicated by the white arrow. (iv) Differential extinction spectrum $\Delta Q/Q$ calculated from the magnetically induced dichroism effect. Contributions from the electric and magnetic dipoles are marked by the dash-dotted and dashed lines, respectively. The overall spectrum (black solid line) exhibits a derivative

spectral shape at the resonances, as indicated by the light blue and light red areas under the curve.[71] c) (i) Schematic representation of the proposed architecture, comprising alternating metal/dielectric (Ag/BIG) layers on a SiO$_2$ substrate. A periodic one-dimensional Au grating is used to couple the incident light to the hyperbolic bulk modes in the multilayer structure. (ii) Real part of the effective permittivity ε$_∥$ and ε$_⊥$, calculated for an infinite multilayer structure. d) $R_{pp}$(+M) and TMOKE for Λ = 830 nm versus incidence angle. The inset shows the normalized magnetic field component of the hyperbolic bulk mode. e) TMOKE versus incident angle at various $n_{an}$. (b) Angle corresponding to the minimum peak of TMOKE as a function of $n_{an}$ represented by solid circles. The solid line represents the linear fitting, whose slope gives $S$ = 75° RIU$^{-1}$. (c) FoM as a function of the angle associated with the minimum peak of TMOKE.[101]

Very recently, it was indeed reported that NM HMM nanoantennas made of Au/TiO$_2$ multilayers can display an enhanced MO response driven by the hyperbolic dispersion via the coupling of either electric or magnetic dipolar optical modes with static magnetic fields (see Figure 4a,b). In particular, the magnetic dipole exhibits a much sharper resonance, thus a much higher Q compared to the those of the electric dipole, and it is almost a non-radiative mode. If we couple this mode with the multipolar mode of a nanoring, although the geometry and fabrication process will be a bit more complex, the results obtained in Ref. [58] would be further amplified, thus potentially approaching, and even exceeding, the theoretical $Q^2$ limit, which considers the bulk dielectric properties of the metal used and not the effective dielectric properties of the system, and perhaps unlock unforeseen MO effects. Finally, a remarkable example of application of these materials has been recently reported by Díaz-Valencia *et al.*,[101] who predicted that MO HMMs can be potentially much more sensitive to refractive index variation compared to common HMMs-based sensors, which can be still considered unbeaten (see Figure 4c-e).[107] This performance can be even improved if proper design rules to increase the quality factor of the modes are implemented.[106] Overall, even if the field of magnetoplasmonics is quite well established, based on these last developments we feel that we are just scratching its surface.

**Ultrafast opto-magnetism with plasmonics**
An interesting emerging aspect in the field of magnetoplasmonics is the possibility to exploit MO phenomena on the nanoscale by using femtosecond (fs) light pulses. For instance, time-resolved MO spectroscopy makes it possible to probe spin dynamics at the timescale of fundamental magnetic interactions, that is the exchange and the spin-orbit coupling (SOC) interactions, responsible for the magnetic order at the macroscale. Furthermore, the reciprocal interaction between spins and fs light pulses is one of the main topics in fundamental magnetism.[108] In particular, the subwavelength regime of light-spin interaction revealed the plasmon-mediated enhancement and emergence of new MO phenomena at ultrafast timescales.[109,110] Novel possibilities of optical control of magnetism on the nanometer scale via plasmonic resonances have emerged as well.

After the discovery of fs demagnetization by ultrashort laser pulses in Ni films,[111] the ability to manipulate optical pulses at a timescale where we can achieve coherent interaction between photons, charges, and spins, has opened the door to exploit the electric-field oscillations of light to control magnetism in unprecedented and unexpected ways.[112–115] Ground-breaking experiments have shown that we can exploit the helicity (right or left circular polarization state) of light to act coherently on the spin moment of electrons,[116–119] as well as induce transient MO effects in nonmagnetic media, where the instantaneous magnetization of the medium results from the light helicity-induced angular momentum of electrons.[120,121] The observation of these effects fueled intense debates over which physical mechanisms drive the processes underlying the electron spin/charge-light helicity coupling, and several theories have been developed, from classical plasma models,[122,123] up to quantum,[124,125] ab initio[126] and relativistic[127] descriptions.

All these groundbreaking experimental and theoretical studies on ultrafast opto-magnetic effects led to the development of the all-optical switching (AOS) of magnetization [128]. The first AOS was demonstrated in a ferrimagnetic amorphous GdFeCo by a single 40 femtosecond circularly polarized laser pulse. After careful examination, it turned out that the magnetic circular dichroism (MCD) and the different demagnetization processes between Gd and Fe sublattices enable this helicity dependent ultrafast (on the order of 30 ps) magnetization control [129–132].

The studies have been evolving mainly using ferrimagnetic and ferromagnetic metallic systems such as GdFeCo, TbFeCo, Co/Pt, Fe/Pt, etc.[118,119,133,134] Although there are some differences in the opto-magnetization processes, every switching mechanism in the metallic systems can be attributed to laser-induced heating near the Curie temperature[132,135] that incurs joule heat loss.

A non-thermal AOS (i.e., AOS not dependent on laser-induced material heating) has been only demonstrated in a dielectric of Co-doped yttrium iron garnet (YIG:Co) through photo-induced magnetic anisotropy and showed ultrafast write-read time (< 20 ps) together with unprecedentedly low heat load (< 6 J/cm$^3$).[136] The Co dopants exhibit strong single ion anisotropy depending on the valence of the ion, which can be modified by light-induced electronic transitions at the Co ions in nonequivalent crystal sites, and thus the magnetic anisotropy of the YIG:Co is controllable with a linearly polarized ultrashort pulsed laser source. Therefore, its AOS can be determined by the orientation of the linearly polarized light due to the crystalline symmetry.[137] The smallest domain size achieved in film samples of the studies above was approximately 5 μm. To meet the storage density over 1 Tbit/inch2, the AOS will have to be achieved beyond the diffraction limit with the lowest possible energy. If the switching is achieved within a 20 × 20 × 10 nm$^3$ domain, the AOS only needs 10 fJ in GdFeCo[138] (see Fig. 4a) and 3 aJ in YIG:Co,[136] approaching the minimum energy required for a stable bit at room temperature, 60 kBT (0.25 aJ). This challenge has been attempted by either miniaturization of opto-magnetic materials by lithography[119,133,138,139] or using plasmonic antennas.[12,109,134] A nanostructured GdFeCo demonstrated a 200 nm domain AOS with a single fs light pulse.[138,139] A few hundred nm long Au wires shaping similar to a dipole antenna on TbFeCo film successfully concentrate a pulsed laser with a wavelength of 1030 nm and the AOS was achieved within ~50 nm in diameter (see Figure 5a).[134] Very recently, magnetoplasmonic crystal arrays composed of Au truncated cones topped with TbCo nanodisks on a glass substrate have exhibited plasmon-assisted thermal demagnetization.[57] The authors not only demonstrated a 3-fold enhancement in the demagnetization efficiency at the localized plasmon resonance, but also showed its mitigation by using the SLR of the array (Figure 5b). Other strategies to achieve more efficient magnetization switching processes by exploiting plasmon-induced thermal effects were also recently reported.[140] Similar thermal effects can also strongly modulate MO effects (e.g., transverse MO Kerr effect),[141] as well as impact demagnetization processes on the ps and nm scales.[142]

The plasmonic enhancement of photo-induced spin precession was reported for YIG:Co with a 1D Au grating. From the comparison with a bare YIG:Co, the amplification efficiency was estimated to be 6-fold at the resonance with the field confinement of 300 nm in the film (Figure 5c).[143,144] A similar enhancement is also seen in the inverse Faraday effect in a Gd-Yb-doped bismuth iron garnet (GdYbBIG) covered with a 1D Au grating and the excitation efficiency of sub-THz spin precession was resonantly enhanced by two orders of magnitude at the resonance within 100 nm deep in the film.[145]

The main function of the plasmon structures in the above applications is to concentrate a laser light. However, it is noteworthy that the same Au/YIG:Co structure as in Ref.[144] exhibits an abrupt phase reversal of the magnetization precession at the plasmon resonance.[146] Since the photo-induced magnetic anisotropy is described by orthogonal electric field components,[136] this is presumed to be the result of the phase shifts in the EM field at the plasmon resonance and the interference between the plasmonic field and the incident field. Although the azimuthal angle dependence of the phase shift in photo-induced spin precession has been reported for a bare GdYbBIG and explained by both the inverse Cotton-Mouton effect and photo-induced magnetic anisotropy,[148] the one caused by plasmonic resonance is not fully figured out yet. Therefore, the finding has revealed another nontrivial role of plasmons in opto-magnetism and suggests new routes to spin manipulation by light with plasmonics.

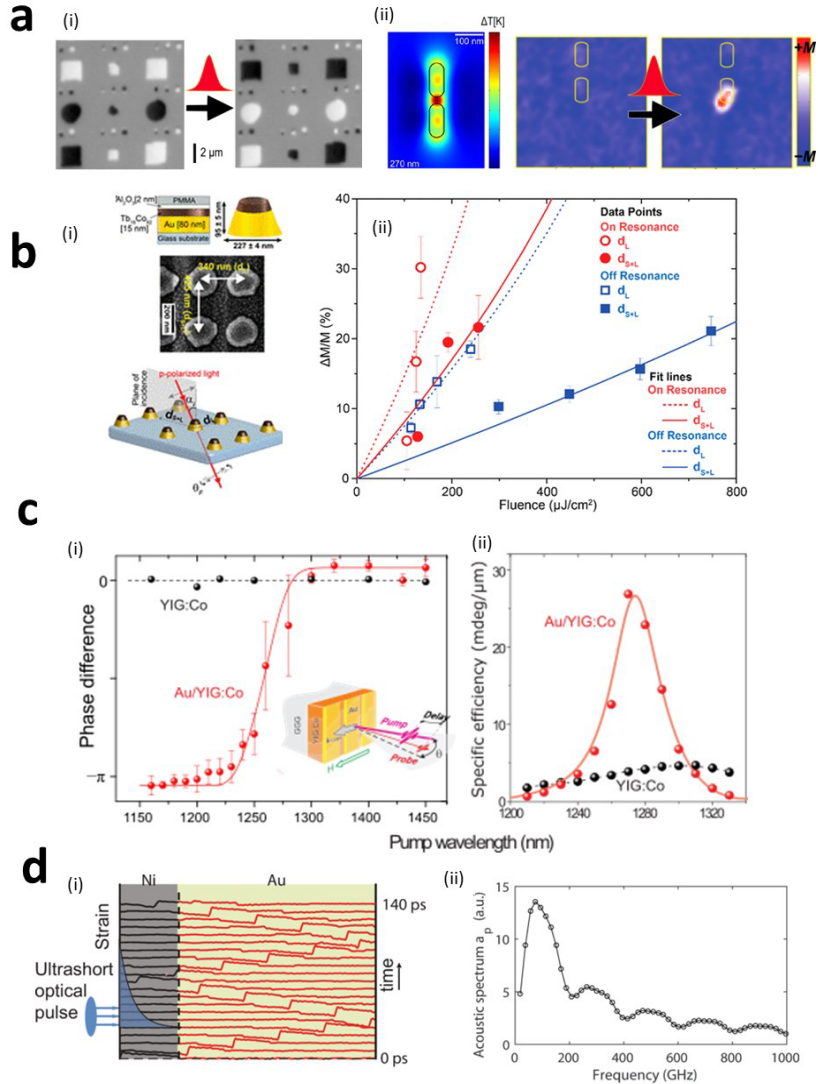

**Figure 5. Nanoscale control of magnetization dynamics with magnetoplasmonics.** a) (i) Magnetization switching by a single femtosecond laser pulse demonstrated in out-of-plane domains with sizes down to 200 nm in GdFeCo nanostructures.[139] (ii) Left: FDTD simulation of the temperature increase in the TbFeCo layer beneath the Au nanoantenna assuming a light pulse of 3.7 mJcm$^{-2}$. Right: Magnetic contrast images before and after a single laser pulse.[134] b) (i, top) Schematic of the nanoantenna structure and composition, (i, middle) SEM of the magnetophotonic surface crystal with two periodicities along the perpendicular directions labeled $d_{S+L}$ and $d_L$, and (i, bottom) experimental geometry for optical transmittance and magneto-optical spectra measurements indicating the plane of incidence and angle of incidence ($\alpha_i$) relative to the surface normal. (ii) Peak values of demagnetization plotted as a function of applied fluence. The solid data points are for measurement along the $d_{S+L}$ direction, and the open data points for measurement along the $d_L$ direction. Solid lines are Curie-law fits for data points along $d_{S+L}$ direction, and dashed lines are Curie-law fits for data points along the $d_L$ direction. Red data points/lines denote on-resonance pumping and blue data points/lines off-resonance pumping of the localized plasmon mode.[57] (c) In the Au/YIG:Co, the magnetic precession reversal (i) is seen across the plasmon resonance at ~1270 nm (i), where the photomagnetic efficiency is enhanced (ii) compared with the bare YIG:Co.[144,146] d) (i) Ultrafast optical excitation of a free-standing Ni/Au bilayer generates ultrashort, picosecond strain pulses $\varepsilon_{zz}(z,t)$ bouncing back and forth in the layer. (ii) The acoustic spectrum for a 30 nm Ni/70 nm Au bilayer obtained using the matrix inversion protocol.[147]

All the above considerations were built upon exploiting only the helicity (right or left circular polarization) of light. An optical beam can carry also orbital angular momentum (OAM).[149,150] Combining OAM with plasmonics, sub-fs dynamics of OAM can be realized in nanoplasmonic vortices,[151–153] which can be confined to deep subwavelength spatial dimension and could offer an excellent time resolution. Thus, the OAM is expected to enter soon the developing area of magnetophotonics, where it could offer a new tool to control magnetism.[14,153] Recently, observations of an interaction between magnetism and an OAM vortex beams were reported.[154,155] By using a hydrodynamic model of the conduction electron gas Karakhanyan *et. al* quantified numerically the relative contributions of helicity and OAM of light to the inverse Faraday effect (IFE) in a thin gold film illuminated by different focused circularly polarized beams carrying OAM.[156] The OAM of light provides a new degree of freedom to control the IFE and the resulting opto-magnetic field has the potential to influence numerous research fields, such as all-optical magnetization switching and spin-wave excitation. These findings support a mechanism of coherent transfer of angular momentum from the optical field to the electron gas, and they pave the way also for new strategies for optical isolation that do not require externally applied magnetic fields. It can already be perceived that exploiting the OAM of photonic beams will become paramount for the future development of magnetoplasmonics.

Overall, plasmonics can make ultrafast opto-magnetism at the nanoscale more fascinating for probing and controlling the dynamics of the electronic and magnetic response. The unprecedented speed and low dissipations of the AOS together with plasmonic localization can be utilized to accommodate ever increasing demand for data storage. It is worth adding that for the miniaturization of thermally stable magnetic domains, not only the localization of the photonic field but also the development of opto-magnetic materials with relatively high anisotropy is important. Although the thermal ultrafast opto-magnetic effects are quite well understood, extensive efforts are currently devoted to broadening experimental and theoretical studies well into the non-thermal regime (< 10 fs, corresponding to the typical plasmon dephasing timescale), to shed more light onto the origin of fundamental opto-magnetic effects. Here, it is not clear yet what is the effect of having a nonthermalized electron population on the charge-spin interaction. In this framework, the study of plasmon dephasing effects on the spin dynamics in MO nanomaterials is still an unexplored terrain. This opportunity can open excellent perspectives in both fundamental and applied aspects of ultrafast magnetoplasmonics. For instance, by working in the sub-100 fs timescale (nonthermal regime), we can control spin dynamics also with metallic nanostructures, thus overcoming the intrinsic limitations placed by ohmic and other losses and maybe reaching higher Q-factor on the fs timescale, opening excellent opportunities towards plasmon-driven MO and magnetic functionalities with a working bandwidth of hundreds of THz.[14]

Finally, it is worth mentioning that, without any need to use plasmonic structures, it is already possible to use fs light pulses to generate magnons with nm wavelengths.[109] An interesting perspective in this direction is offered by the exploitation of artificial freestanding metal-ferromagnet multilayer magneto-phononic cavities to generate a strong phonon-magnon coupling, to excite perpendicular standing spin waves exchange magnons up to 1THz (see Figure 5d).[147] Moreover, if we think about the emerging possibilities enabled by plasmonic picocavities,[157] where the EM field oscillations can be controlled on the pm-scale, we can foresee that in the near future it might be possible to exploit these pico-cavities to excite "atomic-scale" plasmons[158] to control magnetism at the atomic scale with extreme (sub-nm) resolution. By leveraging also on single electron control with plasmonic antennas,[159,160] it will be even possible to unlock the door for driving the spin of single electrons with sub-fs resolution. Ultrafast nonthermal magnetoplasmonics represent an intriguing pathway to either excite ultrafast spin dynamics or understand the magnetic properties of nanostructures at the fs and nm scales, and we foresee that this is just the beginning of a new chapter in the fascinating story of magnetoplasmonics.

## Acknowledgements

N.M. acknowledges support from the Swedish Research Council (grant n. 2021-05784), Kempestiftelserna (grant n. JCK-3122), the European Innovation Council (grant n. 101046920 'iSenseDNA'), the Luxembourg National Research Fund (grant n. C19/MS/13624497 'ULTRON') and the European Commission (grant n. 964363

'ProID'). P.V. acknowledges support from the Spanish Ministry of Science, Innovation and Universities under the Maria de Maeztu Units of Excellence Programme (CEX2020-001038-M) and the project RTI2018-094881-B-I00 (MICINN/FEDER). T.K. acknowledges support via the European Union's Horizon 2020 research and innovation programme under the Marie Słodowska Curie Grant Agreement No. 101029928 (MANACOLIPO). A. G. and F. P. acknowledge the support of Italian Ministry of Education and Research (MIUR) through PRIN Project 2017CR5WCH Q-CHISS.